# Instant square lattice structured illumination microscopy: an optimal strategy towards photon-saving and real-time super-resolution observation

Tianyu Zhao[1,3], Zhaojun Wang[1,3], Manming Shu[1], Jingxiang Zhang[1], Yansheng Liang[1], Shaowei Wang[1], Ming Lei[1,2,*]

[1] MOE Key Laboratory for Nonequilibrium Synthesis and Modulation of Condensed Matter, Shaanxi Province Key Laboratory of Quantum Information and Quantum Optoelectronic Devices, School of Physics, Xi'an Jiaotong University, 710049, China

[2] State Key Laboratory of Electrical Insulation and Power Equipment, Xi'an Jiaotong University, Xi'an 710049, China

[3] These authors contributed equally to this work

*Correspondence: ming.lei@mail.xjtu.edu.cn

**Abstract:**
　　Over the past decade, structured illumination microscopy (SIM) has found its niche in super-resolution (SR) microscopy due to its fast imaging speed and low excitation intensity. However, due to the significantly higher light dose compared to wide-field microscopy and the time-consuming post-processing procedures, long-term, real-time, super-resolution observation of living cells is still out of reach for most SIM setups, which inevitably limits its routine use by cell biologists. Here, we describe square lattice SIM (SL-SIM) for long-duration live cell imaging by using the square lattice optical field as illumination, which allows continuous super-resolved observation over long periods of time. In addition, by extending the previous joint spatial-frequency reconstruction concept to SL-SIM, a high-speed reconstruction strategy is validated in the GPU environment, whose reconstruction time is even shorter than image acquisition time, thus enabling real-time observation. We have demonstrated the potential of SL-SIM on various biological applications, ranging from microtubule cytoskeleton dynamics to the interactions of mitochondrial cristae and DNAs in COS7 cells. The inherent lower light dose and user-friendly workflow of the SL-SIM could help make long-duration, real-time and super-resolved observations accessible to biological laboratories.

**Introduction:**

Understanding how cells function requires a detailed knowledge of their structural organization and of the dynamic interplay of their many constituents, often over long periods of time. However, imaging subcellular structures required an improvement in the resolution beyond the diffraction barrier that limits the best wide-field optical microscopes [1-4]. Super-resolution structured illumination microscopy (SR-SIM), which possesses the merits of high imaging speed and high label compatibility has attracted interests for observing living cells in cell biology[5]. To adapt to the subcellular dynamics in living cells, many new strategies, devices, and algorithms have recently been introduced to SR-SIM to further enhance its imaging capabilities, including temporal resolution, imaging depth, and data fidelity[6,7]. For instance, Huang et al. developed a super-sensitive deconvolution algorithm based on Hessian matrixes to suppress the noise and artifacts in fast SR-SIM, which made it possible to follow the rapid dynamics of moving vesicles in the endoplasmic reticulum at a spatiotemporal resolution of 188 Hz and 88 nm[8]. Also, Guo et al. proposed grazing incidence structured illumination microscopy (GI-SIM) to image dynamic events near the basal cell cortex over thousands of time points at 97 nm resolution and 266 Hz by interleaved reconstruction[9]. In addition, adaptive optics technologies have also been incorporated into SR-SIM, greatly expanding the application of SR-SIM in thick specimens[10]. With the ultrafast imaging speed and the considerable penetration depth, SR-SIM is gaining widespread applications in the investigation of the dynamic behavior of organelles, biomolecules and their assemblies.

Besides the above-mentioned capabilities, phototoxicity of SR-SIM is another critical factor to be considered in applications requiring long-term observations, as it dominates the maximum number of usable consecutive images[11,12]. Conventionally, SR-SIM requires nine raw images to obtain a single super-resolution image, resulting in significantly higher phototoxicity than conventional widefield microscopy. Therefore, reducing the number of raw images is a straightforward solution to increase the observable time points of the SR-SIM[13-15], thereby reducing the photon dose imposed on the sample by nearly half. Unfortunately, the image reconstructions of these methods are mostly based on ill-conditioned problems, leading to degraded spatial resolution and abundant stripe artifacts[16].

In addition to the above methods, exposure time reduction has been shown to be an effective and widely used approach to mitigate photobleaching and phototoxicity effects. Nevertheless, such improvement will easily hit a universe bottleneck owing to the fact that the excessively short exposures will lead to great amount of noise and artifacts in the reconstructed SR images[17]. Fortunately, the super-sensitive deconvolution algorithm proposed by Huang et al[8] enables SR-SIM to work with ultra-short exposure time down to sub-millisecond, which greatly reduces the phototoxity and photobleaching caused by each SIM acquisition. The deep learning method also allows us to directly transform diffraction-limited input images into super-resolved images, which dramatically reduces the light dose imposed on the sample [18]. However, the deep learning-based methods only work well with previously trained specimens, and the potential artifacts contained in the generated SR images cannot be intuitively explained, which severely limits their applications in many fields.

Notably, none of the above solutions took into account the fact that the excessive photon dose imposed on the specimen by SR-SIM is partially due to the redundant illumination caused by the one-dimensional periodic sinusoidal patterns. Specifically, the sample parallel to the stripe pattern is illuminated multiple times in each pattern orientation, which does not provide any additional superresolved information. To address this issue, we develop a new SR-SIM modality by using two-dimensional square lattice structured illumination, which reduces the photon dose imposed on the sample to half. In addition, we developed an ultrafast reconstruction scheme by extending our previous Joint Spatial Frequency Reconstruction (JSFR) algorithm to the SL-SIM[19]. Assisted by the GPU acceleration technique, the specially designed reconstruction algorithm enables an instant "acquisition, reconstruction, and displaying" workflow for real-time SR observation. Finally, we present an optimized SR-SIM strategy for real-time, long-term observation of subcellular structures. The performance of this setup is verified by observing the dynamic interactions of the mitochondrial cristae and the DNAs in COS7 cells. This new technique is anticipated to provide a powerful and user-friendly tool for long-term observation of the rapid interaction of subcellular structures in living cells.

## Methods
### Ultrafast reconstruction algorithm for SL-SIM

In our previous work[19], we proposed the JSFR algorithm to directly attained the SR image by linear superposition of the patterned illuminated raw images with appropriate weighted coefficients in real space, which avoids the complex procedures in frequency domain. Here, we further extended the concept of spatial domain processing for the proposed square lattice structured illumination, which enables us to obtain the super-resolution image in a simple and rapid manner.

For linear SIM system, we first consider the one-dimensional case, in which the illumination field is shifted several steps along x-coordinate. Let $\delta_j$ represent the shifting displacement of step $j$, the acquired image $D_j(x)$ can be described as:

$$D_j(x) = \int O(x')I(x'-\delta_j)H(x-x')\,dx' \tag{1}$$

where $O$, $I$ and $H$ respectively denote the fluorescent emitter distribution of the object, the illumination intensity and the point spread function of the microscope. Assuming the super-resolution image $R(x)$ can be attained by linear superposition of the acquired raw images multiplied with corresponding coefficient function $c_j^{1D}(x)$, that is,

$$R(x) = \sum_{j=1}^{n} c_j^{1D}(x)D_j(x) = \int O(x')[\sum_{j=1}^{n} c_j^{1D}(x)I(x'-\delta_j)]H(x-x')\,dx' \tag{2}$$

For fringe illumination, the illumination intensity distribution can be expressed as

$$I(x) = I_0\{1 + m\cos[2\pi k_0 x + \varphi_0)]\} \tag{3}$$

where, $I_0$, $m$, $k_0$ and $\varphi_0$ denote the mean intensity, modulation depth, spatial frequency and initial phase of the illumination cosine fringe, respectively. We have proved that the super-resolution image can be attained by superposition of the designed phase-shifted patterned illumination images $D_j(x)$ multiplied with each corresponding coefficient function $c_j^{1D}(x)$ as

$R_{SDR} = \sum_{j=1}^{n} c_j^{1D}(x)D_j(x)$, and the function of $c_j^{1D}(x)$ is:

$$\begin{aligned}c_1^{1D}(x) &= \frac{1}{2(1-\cos\Delta\varphi)} + \frac{1-2\cos\Delta\varphi}{2(1-\cos\Delta\varphi)}m\cos(2\pi k_0 x+\varphi_0) - \frac{2\cos\Delta\varphi+1}{2\sin\Delta\varphi}m\sin(2\pi k_0 x+\varphi_0)\\ c_2^{1D}(x) &= -\frac{\cos\Delta\varphi}{1-\cos\Delta\varphi} + \frac{\cos\Delta\varphi}{1-\cos\Delta\varphi}m\cos(2\pi k_0 x+\varphi_0) + \frac{1+\cos\Delta\varphi}{\sin\Delta\varphi}m\sin(2\pi k_0 x+\varphi_0)\\ c_3^{1D}(x) &= \frac{1}{2(1-\cos\Delta\varphi)} - \frac{1}{2(1-\cos\Delta\varphi)}m\cos(2\pi k_0 x+\varphi_0) - \frac{1}{2\sin\Delta\varphi}m\sin(2\pi k_0 x+\varphi_0)\end{aligned} \tag{4}$$

For two dimensional cases, the illumination field must be shifted towards two orthogonal directions with several steps (indicated by the index $i$ and $j$). Let $\delta_i$ represent the shifting displacement of step $i$ in $x$ direction, $\varepsilon_j$ represents the shifting displacement of step $j$ in $y$ direction, we express the illumination pattern $I_{ij}$ by the product of the first order of orthogonal trigonometric basis in $x$ and $y$ directions (which represents a square lattice illumination):

$$I_{ij}(x'-x_i, y'-y_j) = I_0\{1+m\cos[k_0(x'-x_i)+\varphi_0]\}\bullet\{1+m\cos[k_0(x'-x_j)+\psi_0]\} \quad (5)$$

Then we can prove that $c_{ij}^{2D}(x,y)$ can be derived by the exact value of phase shifts via the below equations:

$$c_{ij}^{2D}(x,y) = c_i^{1D}(x) \bullet c_j^{1D}(y) \quad (i=1,2,3; j=1,2,3) \quad (6)$$

Thus, the reconstructed super-resolution image can be directly attained as followed:

$$R_{SDR}(x,y) = \sum_{i=1}^{3}\sum_{j=1}^{3} c_{ij}^{2D}(x,y) \bullet D_{ij}(x,y) \quad (7)$$

That is to say, after all the illumination parameters of the square lattice field including $m$, $k_0$, $\varphi_0$ and $\psi_0$ are determined, the SR image can be calculated by substituting the coefficient functions in Eq. 6 into Eq. 7.

### *Hardware implementation of the instant SL-SIM*

To demonstrate the real-time and low photobleaching and phototoxicity SR imaging strategy, a multi-color, high numerical aperture (NA, 1.49) structured illumination microscope is constructed (Fig. 1). The square lattice structured illumination is generated by the interference of four circularly polarized coherent beams produced by a spatial light modulator (SLM) and a quarter-wave plate. During the SR acquisition, the illumination field is shifted along two orthogonal directions with three steps, that is, nine illumination patterns (Fig. 1a, 1b) are required for single SR acquisition, which is equivalent to conventional SR-SIM based on fringe patterns. In superior to conventional SR-SIM, this scheme requires no elaborate polarization control during the SR acquisition while providing a considerable modulation depth. At the emission path, we used a sCMOS camera (ORCA-Fusion BT, Hamamatsu) with 95% peak quantum efficiency at visible light to detect the emission fluorescence and designed a time sequence that efficiently synchronizes the switch of the global shutter, the pattern generation of SLM and the camera readout interval.

## Results

*Square lattice structured illumination decreases the photon dosage by half*

In classical SR-SIM, the nine raw images used to reconstruct a single SR image are successively collected with shifted and rotated fringe illumination patterns[20] (typically three steps in three orientations) to achieve a nearly isotropic lateral resolution. Comparing with other illumination patterns such as speckle patterns[21] and multifocal patterns[22], the classical fringe pattern provides the best spatial resolution[23] and temporal resolution[8] in the linear excitation regime. It's worth noting that each fringe pattern only contributes to the high-frequency information perpendicular to the direction of the fringe. Along the direction of the stripe pattern, no SR information can be detected by the objective, although the sample is repeatedly illuminated along the stripe pattern, which inevitably leads to a waste of photon dose imposed on the sample. To circumvent this drawback, we propose the square lattice structured light field to be used as the illumination pattern for SR-SIM, which is able to reduce the photon dose imposed on the sample to half at a similar signal level, allowing a longer duration for live cell imaging (Fig. 2a,b). Such an illumination scheme was first mentioned by R. Heintzmann[24], but it has received little attention over the years because its advantage in fringe illumination has hardly been exploited.

To demonstrate the ability of the proposed approach to reduce the light dose, we first compare the luminance decay of microtubules in fixed EGFP-labeled HeLa cells, as shown in Fig. 2a. A total of 6000 raw images with an exposure time of 10 ms are acquired within 10 minutes for both illumination schemes. To eliminate the effect of fluorescence recovery on the comparison results, the raw images are acquired sequentially without any delay. The initial brightness and spatial resolution of the two approaches are comparable. Apparent differences in bleaching rate are clearly observed after 2 minutes. As the illumination is increased, the intensity of the image continues to decrease. Finally, after 10 minutes, both the two cells are quenched into an extremely low emission state. However, the SR images recovered with square lattice illumination still maintain acceptable quality after rescaling the intensity map, while the image with conventional SR-SIM provides bare information but noise. A quantitative comparison is also provided by monitoring the luminance decay rate of the microtubule at much lower illumination levels, as shown in Figure 2b. Nearly 50% of the EGFP fluorescence was rapidly bleached after imaging for 1500 raw frames under conventional fringe illumination, while for the proposed square lattice illumination, it takes about 3000 raw frames to decrease to the same level. Therefore, the square lattice illumination indeed can indeed significantly improve the photon efficiency and reduce the photon budget of SR-SIM. In addition, to demonstrate the difference between the two schemes for live cell imaging, we also verify their performance by observing the microtubule cytoskeleton in live HeLa cells, as shown in Figure 2c. The brightness of the microtubule cytoskeleton imaged with grid illumination is still at a high level after 14 minutes, while the microtubule imaged with conventional SIM shows a significant attenuation of fluorescence intensity

*Instant reconstruction strategies*

For conventional SIM, SR reconstruction is implemented in the frequency domain, where many spectral operations such as OTF compensation, apodization, and Wiener filtering are applied to improve the resolution and suppress the noise and background in the acquired raw images[25]. Due to the complex post-processing procedures, the classical reconstruction algorithm was proved to be time-consuming (typically taking four to eight seconds at 1024×1024 pixels), which make it difficult to carry out online reconstruction during the observation. To improve the reconstruction speed for SR-SIM, Markwirth et al. construct a video-rate and real-time SR-SIM by developing a GPU-enhanced, network-enabled setup. However, its complex arrangement in both hardware and software makes it difficult to be widely implemented in most laboratories. In our previous work[19], we have proposed a joint space and frequency reconstruction (JSFR-SIM) approach to directly obtain super-resolution images for SR-SIM. Superior to the conventional reconstruction algorithm, the JSFR-SIM images were obtained directly by linearly superimposing the patterned illuminated raw images with specific weighting functions in real space. The weighting functions could be pre-calculated with the estimated illumination parameters. As a result, the reconstruction speed was improved by a factor of 80 compared to conventional algorithms, making SIM an ideal tool for real-time super-resolution imaging.

In this work, we further develop an ultrafast reconstruction algorithm by modifying our previous reconstruction scheme to accommodate to SL-SIM (Fig. 1b). Similar to our previous method, the new algorithm only involves simple pointwise multiplication and summation, which also results in an extremely fast reconstruction speed. Moreover, the weighting functions used in the reconstruction can be precalculated with the illumination parameters and be reused for thousands of successive reconstructions, since they do not vary with the specimen information. As shown in Table 1, the running time for full-frame acquisition is less than 35ms, which is even shorter than the acquisition time (90ms) of the raw images. For a smaller region of interest, the reconstruction speed is also much faster than the acquisition speed. Thus, this method is readily applicable to online reconstruction during acquisition. Based on the fast reconstruction approach, we further construct an instant square lattice SIM framework by utilizing the multithreading technique, in which four child threads are created respectively to realizing data acquisition, instant reconstruction, instant result display and instant result storage. After all the nine raw images of a single SR frame are acquired, the reconstruction thread immediately starts the SR reconstruction. Once the reconstruction process finished, the result display thread and the result saving thread instantly display the result on the screen and save the raw images and the reconstructed images into the solid state drive (SSD). The latency between the acquisition and display is calculated to be less than 40ms for a ROI of 1024×1024 pixels.

*Instant observation of subcellular structures*

The reduced photobleaching effect of SL-SIM on the specimen is essentially proven by recording the dynamics of the mitochondrial sequences at a same frame rate and an identical excitation intensity level (Fig. 3a, 3b). In a typical video of 222 raw frames recorded at 2 Hz, abundant interesting events among different mitochondria can be found, as shown in Fig. 3b-d. For instance, a deep-kiss-type mitochondrial interaction between two mitochondria is clearly recorded in Fig. 3c. At first, the offensive mitochondrion took a slight kiss over another mitochondrion (at a timestamp of 00:29) and then retreats a bit along the opposite direction. After a short pause, the tip of the mitochondrion rapidly and forcefully injects itself into the other mitochondrion. The deep kissing process lasted over 30 seconds and the tip extended nearly 3μm. In addition, several successive kiss-and-run-type mitochondrial interactions were also recorded in Fig. 3d, which lasted for approximately 20 s. The mitochondrial tip extended and retracted several times within 5 s. At the same record, we also observed the mitochondrial fusion phenomenon in the same ROI, as indicated by the red box in Fig. 3d. The above diverse mitochondrial dynamics recorded using SL-SIM creates a new paradigm for high resolution, real-time, and low phototoxicity observation for the sub-cellular structures.

*Optically-sectioned super-resolution observation of microtubule cytoskeleton*

In addition, to exploit the optical sectioning capability of the method, we extend the OTF attenuation approach[26] in traditional SIM reconstruction to our SL-SIM, which adapts the method to three dimensional imaging of organelles in live cells. The time series recording the microtubule cytoskeleton dynamics are presented in Fig. 4. The cytoskeleton extends over a depth range of 9 μm, and the whole volume of $74 \times 74 \times 9$ μm$^3$ is acquired at an axial scanning interval of 0.2 μm with 45 steps. The recorded full series contains 9 time points, each acquired by axial scanning through the specimen. With an exposure time of 30 milliseconds, we achieved a volumetric SIM imaging speed of approximately 0.1 fps. In Fig. 4c and 4d, the maximum-intensity-projection images and their colour-coded images clearly illustrate the super-resolution 3D dynamics of the cytoskeleton at the selected regions.

*Visualizing the dynamic interaction of organelles*

Recently, tracking the interaction of organelles, biomolecules and their assemblies with enhanced resolution has drawn increasing interests in physiological, pathological and pharmacological researches[27-29], which technically requires the super-resolution microscope with multi-channel imaging capabilities. To demonstrate the performance of instant SL-SIM in multi-channel imaging, we recorded the dynamic interaction of mitochondria (stained with MitoTracker Green) and mitochondrial DNA (stained with PHB2-mscarlet) in COS-7 cells, as shown in Fig. 5. The multi-channel time series were recorded at a SR imaging speed of 5 Hz, which revealed some interesting events among the mitochondria and the mitochondrial DNA. For instance, in Fig. 5b, we can clearly have an insight into the dynamic behavior of the mitochondrial DNA during the division process of the mitochondria, which is marked by the yellow arrows. In addition, the tiny movement of the mitochondrial DNA during the little-kiss-type mitochondrial interactions was also probed and tracked in Fig. 5c and 5d, which lasted for approximately 6~12 s. When the mitochondria came close to each other, several short and thin mitochondrial tips stretched out and a handful of mitochondrial DNAs were delivered from the upper mitochondria to the lower one. After the kissing behavior finished, the mitochondrial DNA finally stayed inside the lower mitochondria. At the identical record of the same ROI, a second little kiss between the two mitochondria was observed after nearly 20 seconds, as shown in Fig. 5d. Although the aggregation of mitochondrial DNAs at the mitochondrial tips can be clearly observed during the little-kiss behavior, no visible transfer of mitochondrial DNAs is detected in the recording.

**Discussion**

The main focus of this work was to construct a real-time, low-photon-dosage, super-resolution imaging strategy, aiming at providing a gentle and user-friendly tool for biomedical imaging. In superior to the conventional SR-SIM, the proposed instant SL-SIM reduces the light dosage imposed on the specimen to one half, allowing a much longer duration for real-time observation. Meanwhile, the super-resolution image can be recovered, displayed and stored immediately after the raw image acquisition is completed, providing a much smoother workflow for the biomedical researcher, such as that of the conventional wide-field fluorescence microscope or the laser scanning confocal microscope. Unlike the previous real-time SIM approach with GPU acceleration, the proposed method has a much lower latency between the data acquisition and SR image display, because it saves the time-consuming data transfer procedures from host to host and from host to graphics devices. In addition, the generation of square lattice structured illumination does not require complex polarization control, providing a simple and economical way to construct a SIM setup.

It's worth noting that instant square lattice SIM can also be combined with Hessian-denoising algorithms to further reduce the photon budgets of SR imaging or SR-OS imaging in an offline manner. We can further improve the quality of the SR images with Hessian-denoising algorithms after the real-time observation is finished. Thus, the random noise in the second half of the playback can be effectively suppressed. In addition, square lattice structured illumination could also be introduced into the nonlinear excitation regime to further improve the spatial resolution. In fact, the idea of exploiting the nonlinear excitation capability of four coherent beams has been applied in RESOLFT[30] to improve the field of view and the imaging speed. However, the polarization states of the four coherent beams are different from that of SL-SIM. We believe that the application of four-beam-interference in nonlinear excitation can be further expanded by exploring diverse types of structured fields.

In summary, in comparison to current SR microscopy techniques, the proposed real-time SL-SIM offers an instant, super-resolution imaging method with a reduced photon budget. Under an illumination intensity of half of that required in conventional SIM, SL-SIM enables a real-time observation of subcellular structures in live cells with sub-100 nm lateral resolution with hundreds of time points. Moreover, the validated multichannel observation capability of this approach also provides a promising tool to investigate the interaction of organelles, biomolecules and their assemblies. This new imaging strategy is anticipated to provide a real-time, gentle and super-resolution imaging approach to investigate the dynamics of subcellular structures for physiological, pathological and pharmacological research.

**Acknowledgements**

This work was supported in part by the National Key Research and Development Program of China (Nos. 2023YFF0722600, 2022YFF0712500); Natural Science Foundation of China (NSFC) (Nos. 62135003, 62205267); The Innovation Capability Support Program of Shaanxi (No. 2021TD-57); Natural Science Basic Research Program of Shaanxi (Nos. 2022JZ-34, 2024JC-YBMS494); The Fundamental Research Funds for the Central Universities (xzy012023033)

**Author contributions**

M.L. T.Z. and Z.W. conceived and designed the project. T.Z. and M.S. designed and constructed the imaging system and collected data. Z.W. and J.Z. developed the controlling software. T.Z. and Y.L. built the reconstruction software with CPU and GPU environments. S.W. prepared the biological specimens and provide descriptions about the sample preparation. All authors contributed to results interpretation. M.L. T.Z. and Z.W. drafted the manuscript, to which all authors contributed.

**Competing interests**: The authors declare no competing interests.

# Figures

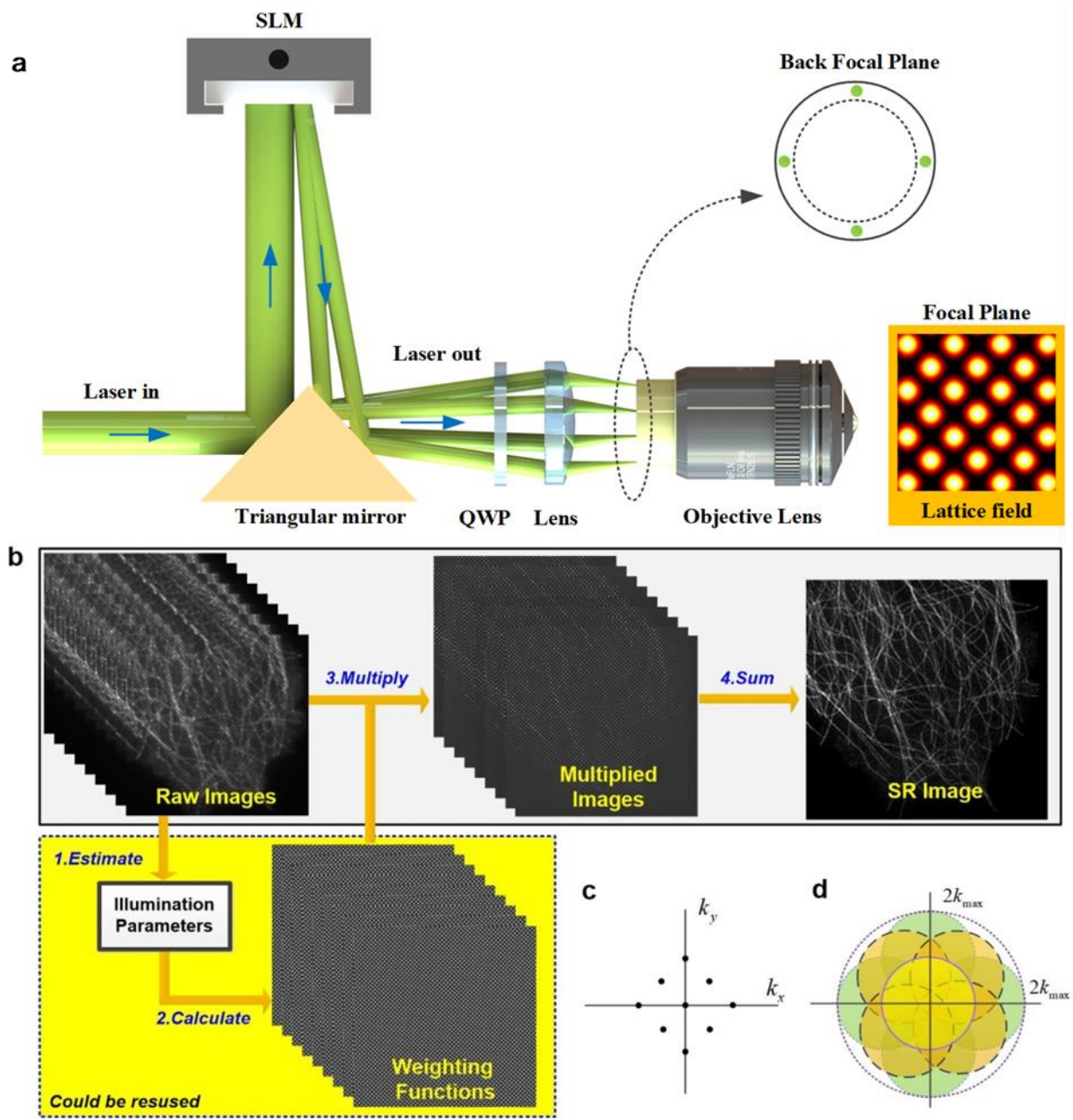

**Figure 1. Diagrams of the hardware and algorithm of SL-SIM.** (a) Schematic diagram of the SL-SIM setup. SLM: spatial light modulator; QWP: quarter wave-plate. (b) Flow chart of the ultrafast reconstruction algorithm, which mainly includes four steps: the precise determination of the illumination parameters, calculation the weighting functions with the estimated parameters, multiplying the weighting functions with the raw images, and obtain the SR image by summing all the multiplication results. (c) The wave vector of the square lattice structured illumination, which is composed of nine frequency spikes. (d) The schematic diagram of the extended spectrum with square lattice structured illumination.

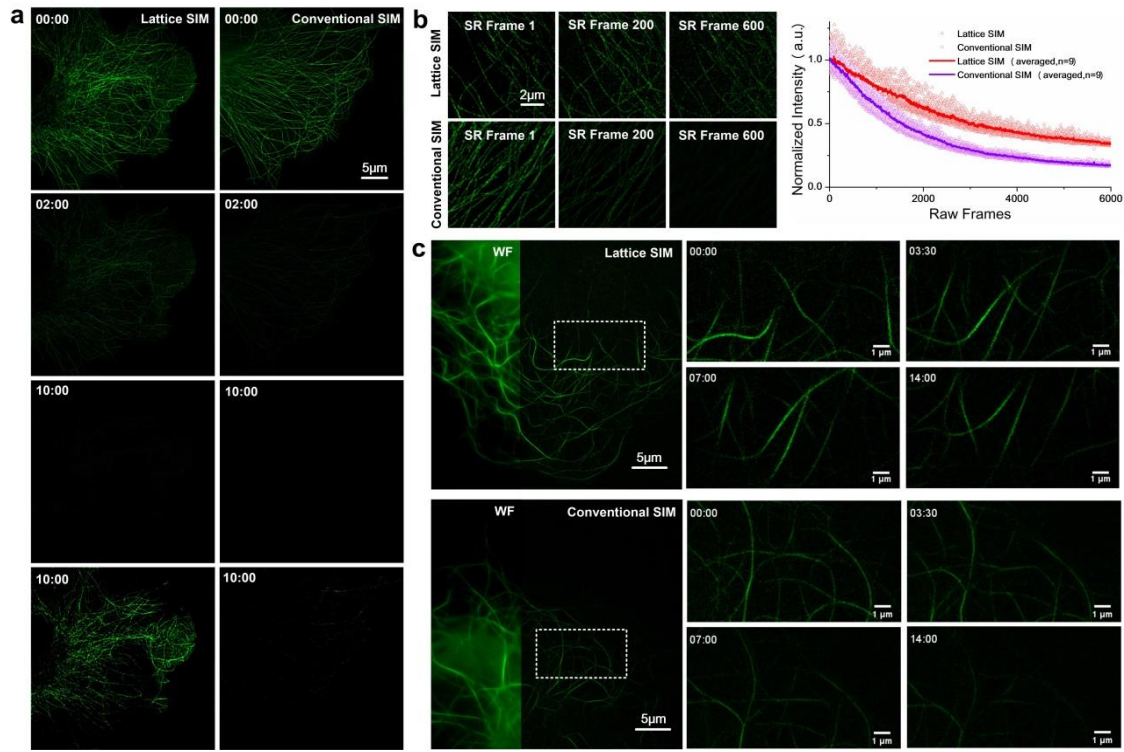

**Figure 2. The square lattice structured illumination increases the frame number for long-term super-resolution observation of microtubule cytoskeleton in HeLa cells.** (a) Consecutive SR imaging of fixed microtubule cytoskeleton to compare the photobleaching speed of conventional SIM and SL-SIM. For the two imaging modalities, the laser power is carefully adjusted to guarantee the specimens are excited at a similar intensity level. In order to provide a fast glimpse over the difference of bleaching speed, we speed up the bleaching process by tuning laser power up to 100%. (b) Comparison of the luminance decay rate of microtubule cytoskeleton in fixed HeLa cell under different imaging modalities. (c) Long term observation results of microtubule cytoskeleton in living HeLa cells with the two imaging approaches. The laser power is set as 10% in order to slow down the photobleaching.

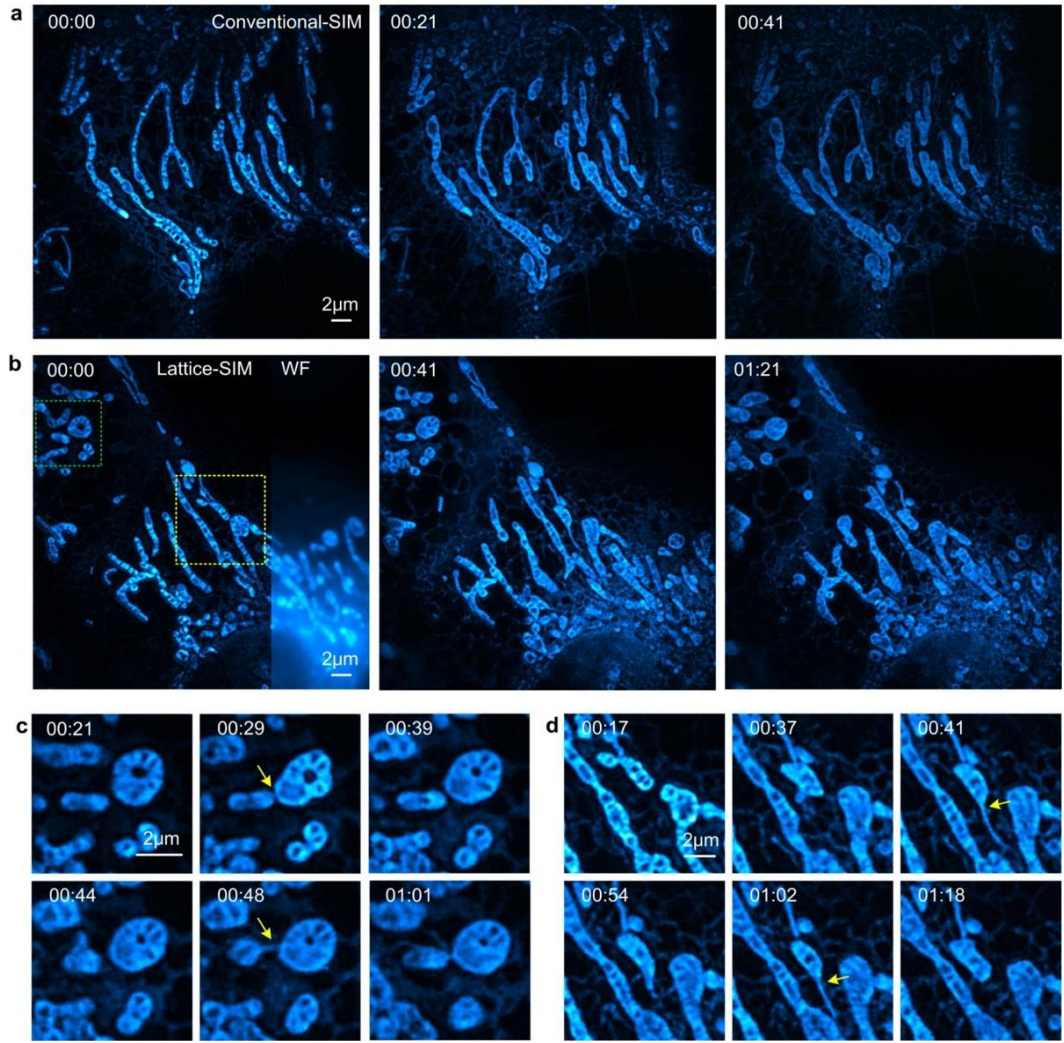

**Figure 3. The square lattice structured illumination enables long time-lapse SR imaging of mitochondrial dynamics without introducing with reduced photobleaching effect.** (a) Time-lapse SR imaging of mitochondria in HRPE cells using conventional SIM. (b) Time-lapse SR imaging of mitochondria in HRPE cells using SL-SIM. Images were acquired with a frame rate of 2 Hz. (c) Close-up view of the dashed green box in (b). (c) Close-up view of the dash yellow box in (b).

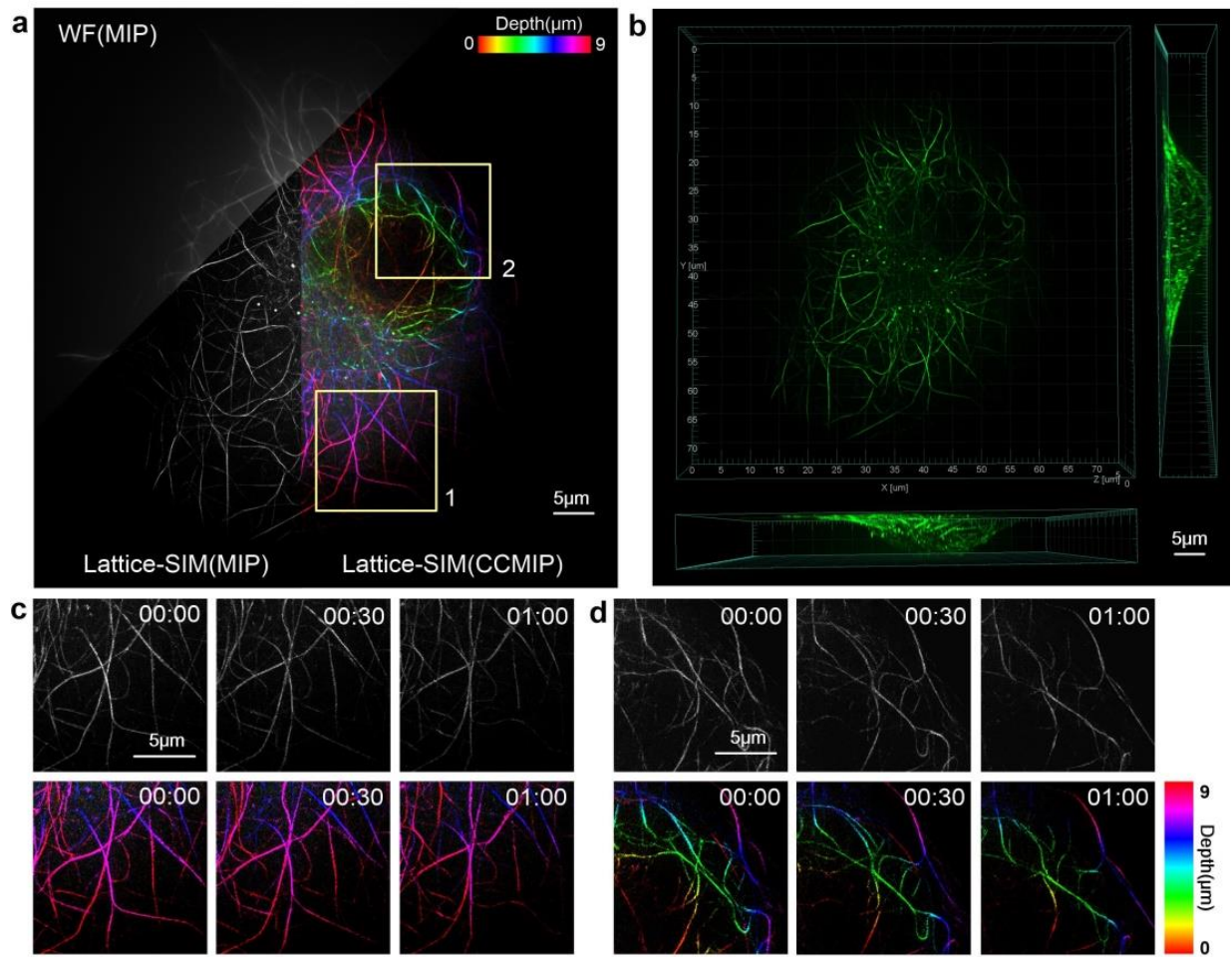

**Figure 4. Time series of SR-OS-SIM data following microtubule cytoskeleton dynamics (stained with EGFP) in COS-7 cells.** (a) The widefield image, the super-resolution maximum-intensity-projection (MIP) image and its colour-coded image (CC-MIP) of the microtubule cytoskeleton (the 1st volume of the time series). (b) The rendered 3D structures of the microtubule cytoskeleton in (a). (c) and (d) are respectively the time series (the first, the fourth and the seventh volume) of the close-up view of the yellow-boxed regions in (a). The upper row is the super-resolution MIP image, while the bottom row illustrates the CC-MIP images of the specimen. The brightness of the series has been renormalized to compensate for photobleaching effect.

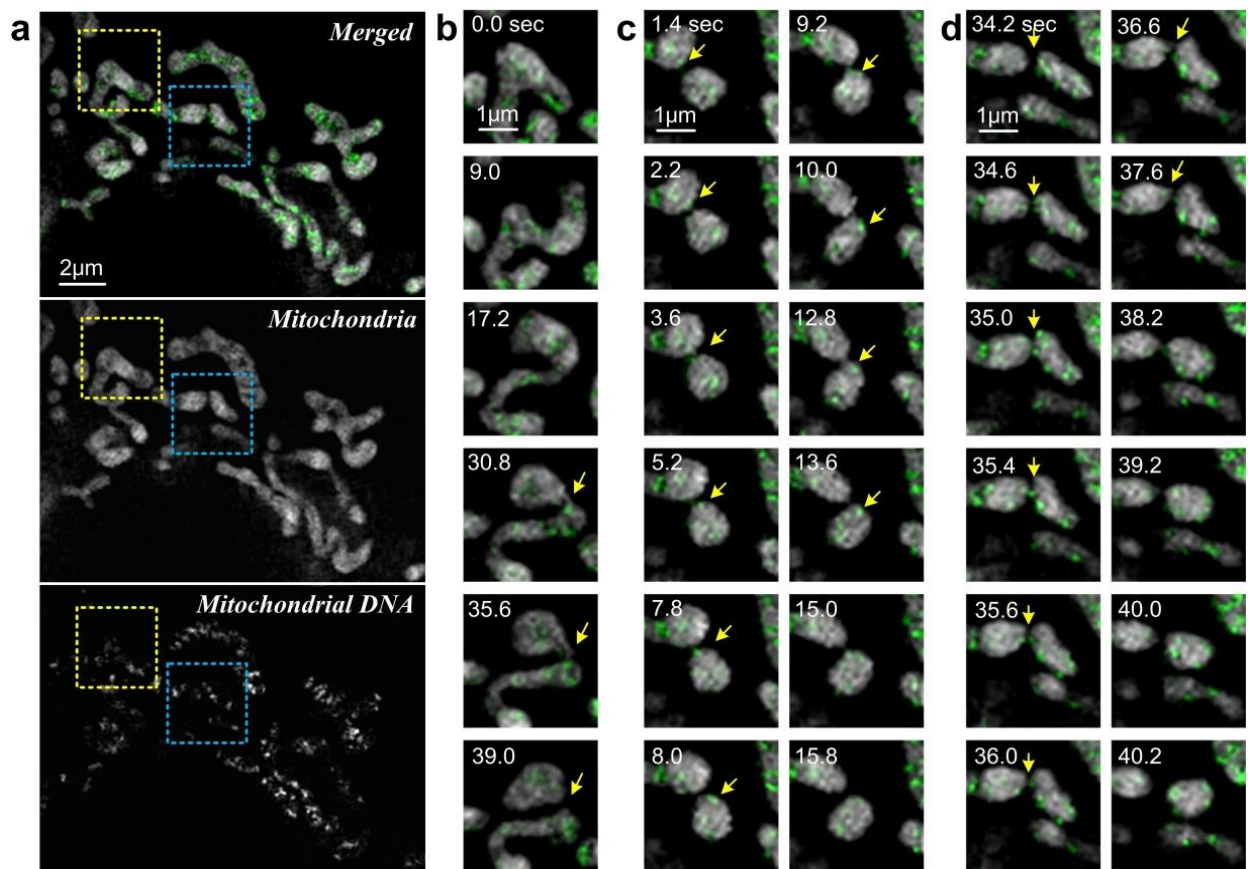

**Figure 5. Time series of SR data recording the dynamic interaction of mitochondria (stained with MitoTracker Green) and mitochondrial DNA (stained with PHB2-mscarlet) in COS-7 cells.** (a) Merged and individual super-resolution images of mitochondria and mitochondrial DNA. (b) Close-up view of the dynamic interaction of mitochondria and mitochondrial DNA at the yellow dash box in (a). (c) and (d) are two different actions recorded by the close-up view of the dynamic interaction of mitochondria and mitochondrial DNA at the blue dash box in (a). The brightness of the series has been renormalized to compensate for photobleaching effect.

Table 1 Execution time of SL-SIM as a function of image size.

| Input image size | Output image size | SL-SIM(ms) | |
|---|---|---|---|
| | | CPU[a] | GPU[b] |
| 1024×1024 | 2048×2048 | 1401.9±15.0 | **32.5±0.8** |
| 512×512 | 1024×1024 | 293.6±4.7 | **7.7±0.7** |
| 256×256 | 512×512 | 73.0±0.7 | **2.9±0.2** |

a. The execution time with indicated image dimensions were executed on a personal computer (Intel Core i7-7900K, 16GB RAM) with MATLAB (R2019a, Math Works Inc.). Before reconstruction, the raw images are up-sampled by a scale of 2 to improve the sampling frequency. The values shown are from 2000 separate processing events of each image, with the times averaged.

b. The GPU code was executed in a mid to high-end graphic card (Nvidia GTX 4070Ti).